\documentclass[conference]{IEEEtran}
\IEEEoverridecommandlockouts
\usepackage{cite}
\usepackage{amsmath,amssymb,amsfonts}
\usepackage{algorithmic}
\usepackage{graphicx}
\usepackage{textcomp}
\usepackage{xcolor}
\def\BibTeX{{\rm B\kern-.05em{\sc i\kern-.025em b}\kern-.08em
    T\kern-.1667em\lower.7ex\hbox{E}\kern-.125emX}}
\usepackage{comment}
\usepackage{amsmath}
\usepackage{breqn}
\usepackage{array}
\usepackage{multirow}
\usepackage{braket}
\usepackage{booktabs}
\usepackage{colortbl}
\usepackage{csquotes}
\usepackage{url}

\begin{document}

\title{ Minimum-length chain embedding for the phase unwrapping problem on D-Wave's Advantage architecture}

\author{

\IEEEauthorblockN{Mohammad Kashfi Haghighi}
\IEEEauthorblockA{
\textit{Department of Electrical and Computer Engineering} \\
\textit{University of Victoria}\\
Victoria, Canada \\
mohammadkashfi@uvic.ca
}
\and

\IEEEauthorblockN{Nikitas Dimopoulos}
\IEEEauthorblockA{
\textit{Department of Electrical and Computer Engineering} \\
\textit{University of Victoria}\\
Victoria, Canada \\
nikitas@ece.uvic.ca
}
}
\maketitle

\begin{abstract}
With the current progress of quantum computing, quantum annealing is being introduced as a powerful method to solve hard computational problems. In this paper, we study the potential capability of quantum annealing in solving the phase unwrapping problem, an instance of hard computational problems. To solve the phase unwrapping problem using quantum annealing, we deploy the D-Wave Advantage machine which is currently the largest available quantum annealer. The structure of this machine, however, is not compatible with our problem graph structure. Consequently, the problem graph needs to be mapped onto the target (Pegasus) graph, and this embedding significantly affects the quality of the results. Based on our experiment and also D-Wave's reports, the lower chain lengths can result in a better performance of quantum annealing. In this paper, we propose a new embedding algorithm that has the lowest possible chain length for embedding the graph of the phase unwrapping problem onto the Pegasus graph. The obtained results using this embedding strongly outperform the results of Auto-embedding provided by D-Wave. Besides the phase unwrapping problem, this embedding can be used to embed any subset of our problem graph to the Pegasus graph.
\end{abstract}


\section{Introduction}
Two-dimensional phase unwrapping is the process of recovering unambiguous phase values from a two-dimensional array of phase values known only modulo $2\pi$ rad. 
The measured phase is also affected by random noise and systematic distortions.
This problem arises when the phase is used as a proxy indicator of a physical quantity, which is the time delay between two signals in the case of interferometric SAR (InSAR) \cite{curlander1991synthetic} and can be used to extract accurate three-dimensional topography. 
As the phase is observable only on a circular space where all measured values are mapped to the range $(-\pi, \pi]$, the observed data must be mapped back to the full range of real phase values to be meaningful.
For unwrapping, the sampling rate is assumed to be suitable to prevent aliasing, i.e., the absolute difference in phase between two adjacent points is assumed to be smaller than $\pi$. 

The most commonly used unwrapping technique is based on network programming strategies that formulate the problem as a minimum cost flow (MCF) \cite{costantini1998novel} problem. 
One of these solvers is the sequential tree-reweighted message passing (TRWS) algorithm \cite{kolmogorov2006convergent}. 
However, since the InSAR images can be quite large---normally larger than $600M$ pixels---the process of phase unwrapping via TRWS can take a prohibitively long time. 
Hence, we explore whether a quantum computing system could be a potential candidate for solving such a problem.

Quantum annealing systems are able to solve problems in quadratic unconstrained binary optimization (QUBO) form. Any unconstrained quadratic integer problem with bounded integer variables can be transformed by a binary expansion into QUBO \cite{glover2018tutorial}. 
The phase unwrapping problem is a quadratic unconstrained problem and it can be mapped to a QUBO.

InSAR images tend to be quite large, often exceeding $600M$ pixels, requiring at minimum 
a 600M-qubit quantum annealer; such a machine is not currently available. 
To overcome present day technology limitations, we have developed a method where we partition the image then use quantum annealing on the individual partitions to obtain suboptimal labelling, and then use quantum annealing in a second phase to obtain labels that approach the ones obtained through classical methods. 
We have named our method “super-pixel decomposition". 

We have reported the preliminary results of the super-pixel decomposition method in \cite{kelany2020quantum},
while in \cite{kelany2022quantum} and \cite{KelanyCCECE2022} we presented enhancements of the proposed method by utilizing additional (marginal) pixels in each of the sub-images,
and refined our experimental analysis by using a larger dataset of synthetic images providing us with statistically more robust results.

In our experimentation, we observed that the mapping of the problem on the annealing machine plays a crucial role in achieving improved performance. This has also been observed by the developers of D-Wave annealers \cite{KingChainLength} stating that there is \enquote{strong evidence that chain length plays an important role in performance}.
Achieving thus symmetrical embeddings with a minimum chain length promise an increased performance.

In this work, we study a variety of embeddings for the phase unwrapping problem on D-Wave's Advantage architecture and experimentally study the impact of the chain length on the performance.   
One of the embeddings we have developed has chain lengths of one meaning that our embedding utilizes the direct links of the underlying Pegasus graph.

The rest of the paper is organized as follows: 
In Section \ref{sec:background} we provide a background of the phase unwrapping problem and the Pegasus graph deployed on D-Wave's Advantage architecture, 
in Section \ref{sec:methodology} we explain our embedding methodology, 
in Section \ref{sec:experiments} we present the experimental results, and 
we conclude with Section \ref{sec:conclusion}.

\section{Background}\label{sec:background}

\subsection{Phase Unwrapping Formulation}
Let $\phi$, $\varphi$, and $k$ denote the unwrapped phase, the wrapped
phase, and an integer label to be estimated. For the phase of a pixel $i$, we have, 
\begin{equation}
\label{eq:1}
\phi_{i}=\varphi_{i}+2\pi k_{i}
\end{equation}

Phase unwrapping, i.e. estimating $\phi$, is an ill-posed problem. The Nyquist criterion \cite{chen2002phase} ensures that phase unwrapping can be performed correctly.
The unwrapping problem can then be expressed as an optimization problem of the cost function \cite{kelany2020quantum},
\small
\begin{equation}
E=\sum_{(s,t)\text{\ensuremath{\in}}A}W_{st}\left(k_{t}-k_{s}-a_{st}\right)^{2}+\sum_{s\text{\ensuremath{\in}}A}\omega_{s}\left(k_{s}-a_{s}\right)^{2}
\label{eq:phase_uwrapping}
\end{equation}
\normalsize
where $k_{i}$ are the labels that will determine the original phase as per \eqref{eq:1}, $A$ is the set of pixels in the SAR image, $W_{st}$ are weights defining the neighbourhood structure and $a_{ij}$ are constants obtained from the image. 
The weights $W_{st}$, $\omega_{s}$, and the bias $a_{s}$, are chosen heuristically and represent ad-hoc information one may have on the scene.

\subsection{Quantum Annealing}

Quantum computational systems, such as the ones developed by D-Wave, use quantum annealing to locate the ground-state of an artificial Ising system \cite{johnson2011quantum}, which is a QUBO problem.
Any quadratic unconstrained optimization problem can be cast as a QUBO problem by expressing the integer variables in binary.

Analytical and numerical evidence indicates that quantum annealing can outperform simulated annealing \cite{heim2015quantum}.

D-Wave Systems provides implementations of different quantum annealing systems, starting from the D-Wave One announced in 2011  \cite{johnson2011quantum}. 
We have used the D-Wave 2000Q\_6 machine ($2041$ qubits) and currently the D-Wave Advantage ($5640$ qubits) machine\cite{boothby2020next}.

An Ising Hamiltonian describes the behaviour of such a system as

\begin{equation}
H_{p}=\sum_{i=1}^{N}h_{i}\sigma_{i}^{z}+\sum_{i,j=1}^{N}J_{ij}\sigma_{i}^{z}\sigma_{j}^{z}
\label{eq:Hamiltonian}
\end{equation}
where $h_{i}$ is the energy bias for spin $i$, $J_{ij}$ is the coupling energy between spins $i$ and $j$, $\sigma_{i}^{z}$ is the Pauli spin matrix, and $N$ is the number of qubits. 
Quantum annealing on this system is achieved by the gradual evolution of the Hamiltonian system\cite{johnson2011quantum},
\begin{equation}
H\left(t\right)=\Gamma\left(t\right)\sum_{i=1}^{N}\Delta_{i}\sigma_{i}^{x}+\varLambda\left(t\right)H_{p}
\end{equation}
As time passes, $\Gamma$ decreases from 1 to 0 while $\varLambda$ increases from 0 to 1.
If the annealing is performed slowly enough, the system stays in the ground state of $H(t)$ for all times, $t$, ending up at the end of the annealing at the ground state of $H_p$.
The Hamiltonian in \eqref{eq:Hamiltonian} can be rewritten in vector form as $H\left(s\right)=s^{T}Js+s^{T}h$, in the form of a QUBO problem\cite{zaribafiyan2017systematic}.

As used in the rest of this paper, the objective function is expressed in QUBO form in scalar notation, and is defined as follows:
\begin{equation}
C\left(x\right)=\sum_{i}a_{i}x_{i}+\sum_{i<j}b_{i,j}x_{i}x_{j}
\end{equation}
where $x\in\left\{ 0;1\right\} ^{n}$ is a vector of binary variables
and $\left\{ a_{i};b_{i;j}\right\} $ are real coefficients. 

Before an application problem can be solved on a quantum annealer, it must first be mapped into QUBO form.
As a first step in transforming the InSAR problem into a QUBO problem, the $k_i$ label that is non-binary valued must be transformed into binary valued. Let $k_{i}\in\left\{ 0,D_{i}-1\right\}$, where $D_{i}$ is
the number of allowed values (labels) for $k_{i}$. This can be achieved
by writing $k_{i}$ in binary. The binary transformation restricts the
number of new-valued binary variables required to represent $k_{i}$.
Let $d_{i}=\lceil \text{log}_{2}D_{i}\rceil$ and $k_{i}=\langle\mathbf{2},\mathbf{x_{i}}\rangle$
where the vector $\mathbf{x}_{\mathbf{i}}=[x_{i,d_{i}},\cdots,x_{i,1},x_{i,0}]$
represents the bits of $k_{i}$ and $\mathbf{2}=[2^{d_{i}},\cdots,2,1]$
is the vector of powers of two. Equation \eqref{eq:phase_uwrapping} can be written in QUBO
form as:
\begin{multline}
E=\sum_{(s,t)\text{\ensuremath{\in}}A}W_{st}\left(\sum_{i}b_{i}x_{i,t}-\sum_{i}b_{i}x_{i,s}-a_{st}\right)^{2}\\
+\sum_{s\text{\ensuremath{\in}}A}\omega_{s}\left(\sum_{i}b_{i}x_{i,s}-a_{s}\right)^{2}
\label{eq:unwrap_binary}
\end{multline}
where $b_{i}$ is the weighting coefficient for the binary variable $x_i$ ($b_{i}=2^i$ in the case of the binary encoding). 

Many problems can be formulated as QUBO to take advantage of quantum annealing, potentially converging faster than other techniques to an optimum solution\cite{jooya2017accelerating}.

\subsection{Chimera Network}\label{subsec:chimera}
\begin{figure}
\centering
  \includegraphics[width=0.6\columnwidth]{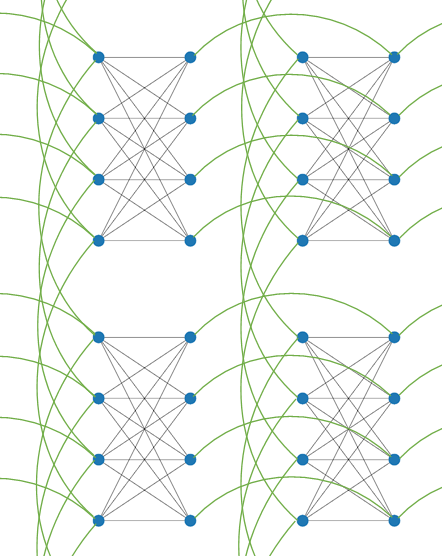} 
  \caption{Chimera unit cells and their connections. Green edges show connections between nodes of different unit cells.}
  \label{fig:chimera}
\end{figure}

The Chimera graph is the underlying architecture of the D-Wave $2000Q\_6$ system.  A Chimera cell consists of 8 qubits located in two columns. Nodes in each column of a Chimera cell are connected to all nodes of the other column but have no connections to nodes within their own column. Figure \ref{fig:chimera} shows Chimera cells together with connections to neighboring cells \cite{dwave2023docs}. In the D-Wave $2000Q\_6$ system, there is a matrix of $16 \times 16$ Chimera unit cells in the whole graph. Unit cells are connected horizontally and vertically to adjacent unit cells. Chimera graph has a degree of 6 as each node is connected to four nodes of their own cells and two nodes of adjacent cells. Figure \ref{fig:cellTypeChimera} shows a more complete representation of the intra-unit-cell connections. 

\subsection{Pegasus Network}\label{subsec:pegasus}
\begin{figure}
\centering
\includegraphics[width=.45\textwidth]{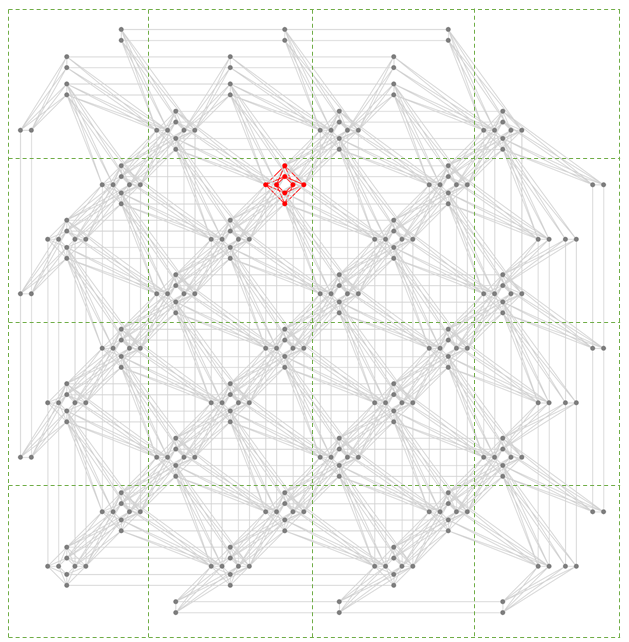}
\caption{Pegaus graph. Unit cells are shown in green squared grids. The red graph is a Chimera unit cell in the Pegasus graph.}
\label{fig:pegasus}
\end{figure}

The Pegasus graph is the underlying architecture of D-Wave's Advantage machine. This architecture provides more qubits and couplers in comparison to D-Wave's previous architecture which deployed Chimera architecture. A section of this architecture is shown in Figure \ref{fig:pegasus}. The Pegasus graph includes the Chimera graph as its sub-graph. Each Pegasus unit cell (shown with square grids in Figure \ref{fig:pegasus}) consists of three Chimera unit cells. The Pegasus graph has a degree of 15 and a nominal degree of 12\cite{boothby2020next}. This architecture also includes $K_4$ (complete graph of degree 4), and also $k_{6,6}$ (bipartite graph of degree 6).

\section{Methodology}\label{sec:methodology}
In this section, we summarize our methodology in mapping our phase unwrapping problem to the Pegasus network natively and also onto the Chimera graph symmetrically.
\subsection{The phase unwrapping problem graph}\label{sec:phaseUnwrappingGraph}
We want to map 2D images with pixels of a maximum label of 3. 
binary variables of these images can be specified using three coordinates, the first two coordinates denote the position of pixels in the vertical and horizontal directions. Specifically, we labeled the vertical coordinate as $i$ and the horizontal coordinate as $j$. The third coordinate, denoted as \enquote{b}, is used to distinguish between the two bits of each pixel. $q=1$ represents the Most Significant Bit (MSB), and $q=0$ represents the Least Significant Bit (LSB). For example, $(i, j, q) = (2, 3, 1)$ represents the MSB of the fourth pixel of the third row in the image as $i$ and $j$ start from 0.
\par
As we are using four-neighbor connectivity for unwrapping, the connections of each pixel with its top, bottom, left, and right pixels should be taken into account. Each pixel has 2 binary variables that need to be connected. Each of these variables is also connected to the binary variables of its 4 adjacent pixels. This results in a total of 9 connections per binary variable: One connection to its own pixel and two connections to each of four neighboring pixels. However, marginal pixels may exhibit fewer connections.
\par
\begin{figure}
\centering
\includegraphics[width=.45\textwidth]{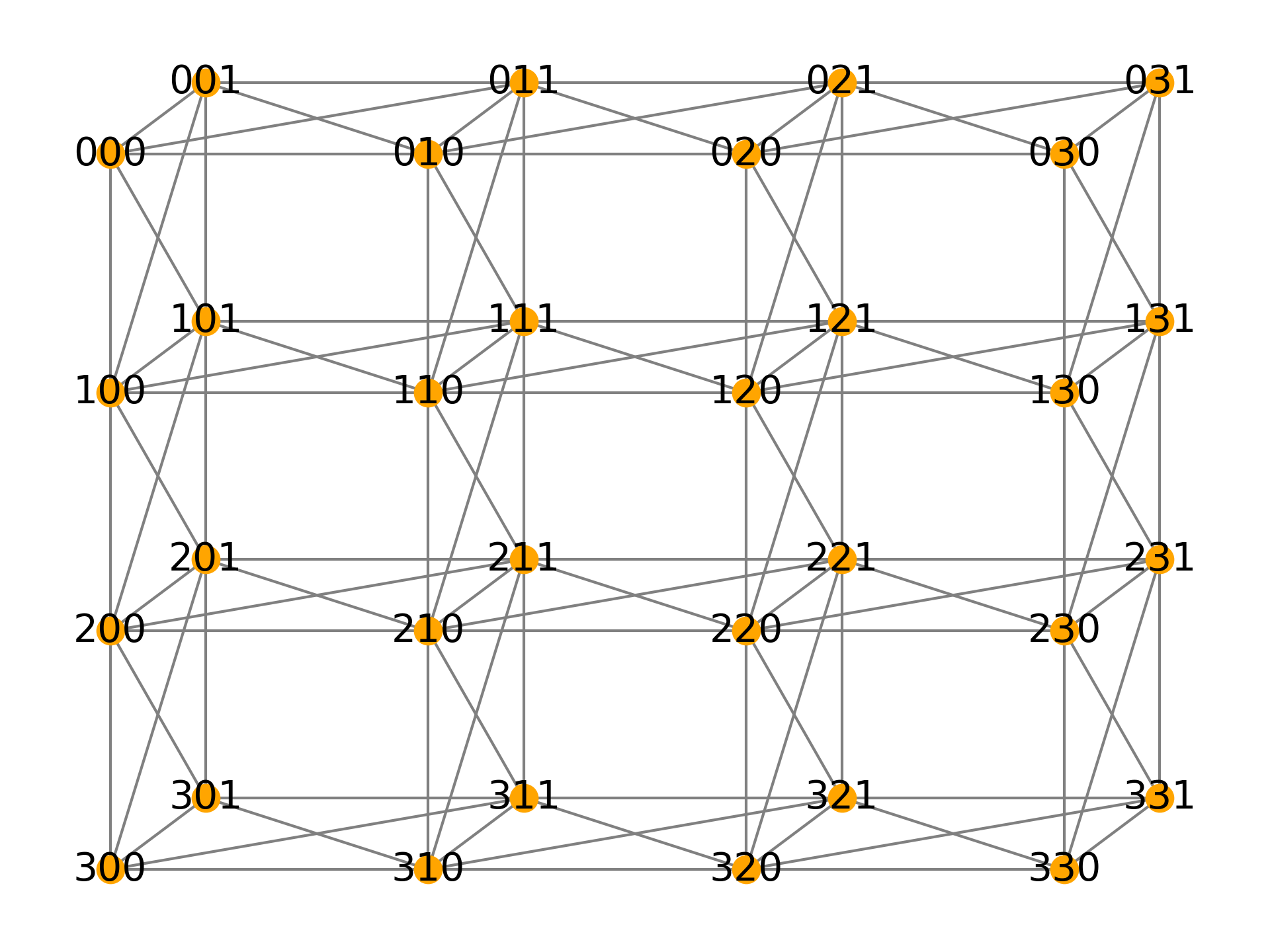}
\caption{The graph of phase unwrapping problem for a $4\times4$ image. Nodes are labeled with a 3-digit number. The first digit (i) corresponds to the column of the pixel, the second digit (j) corresponds to the row of the pixel (both start from $0$), and the third digit (q) specifies which bit of the pixel is being referred to.}
\label{fig:problem graph}
\end{figure}

Figure \ref{fig:problem graph} shows the graph of our problem for an image with the size of $4\times4$. This graph can be expanded to larger sizes while maintaining the same structure.
\par
In the following two sections, we shall focus on the method of mapping the phase unwrapping problem  (and any problem described by the graph in Figure \ref{fig:problem graph}) to two D-Wave architectures namely the Chimera and Pegasus. These architectures deploy the Chimera and Pegasus interconnects discussed earlier. The focus of our approach is to obtain mappings that are as symmetric as possible with minimum-length chains

\subsection{Mapping onto the Chimera graph}\label{sec:mapChimera}

As mentioned before, our problem forms a graph of degree of 9. On the other hand, the degree of the Chimera graph is 6. Consequently, it's not possible to map the binary variables directly to Chimera nodes. Alternatively, we chain multiple qubits in the Chimera graph and map the variables to that qubit chains. This is accomplished symmetrically in \cite{kelany2022quantum}.
\par
\begin{figure}
\centering
\includegraphics[width=.32\textwidth]{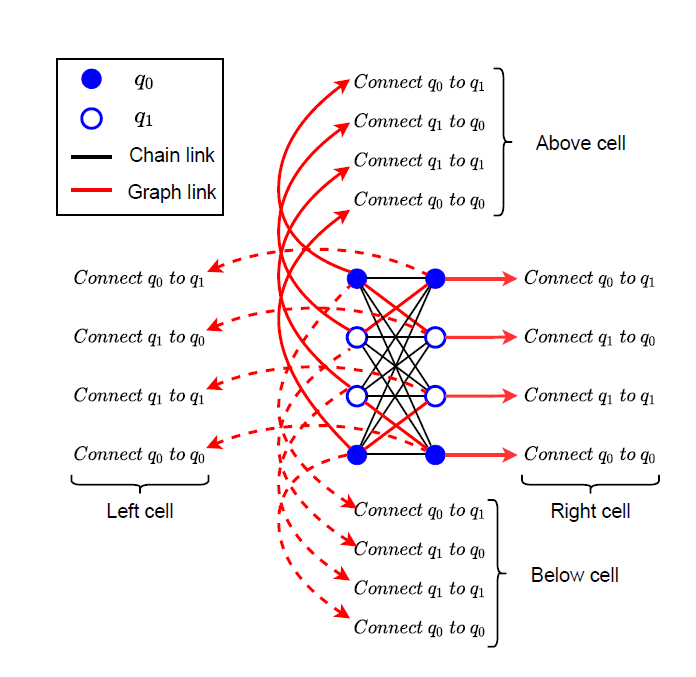}
\caption{Mapping a pixel onto a Chimera unit cell. Each binary variable of a pixel is mapped to a chain of four qubits \cite{kelany2022quantum}.}
\label{fig:cellChimera}
\end{figure}

We mapped each pixel to one Chimera cell. Each binary variable of a pixel is mapped to a chain of four qubits inside a Chimera cell as illustrated in Figure \ref{fig:cellChimera}  \cite{kelany2022quantum}. As qubits need to be connected to each other in a qubit chain, and there aren't any connections between qubits of a single column in a Chimera unit cell, we mapped binary variables into the qubits of two rows in a Chimera cell.  In Figure \ref{fig:cellChimera}, links between qubits of a single binary variable (qubit chain) are shown in red while connections between two qubit chains corresponding to LSB and MSB of a pixel are shown in black \cite{kelany2022quantum}.

\par

\begin{figure}
\centering
\includegraphics[width=.32\textwidth]{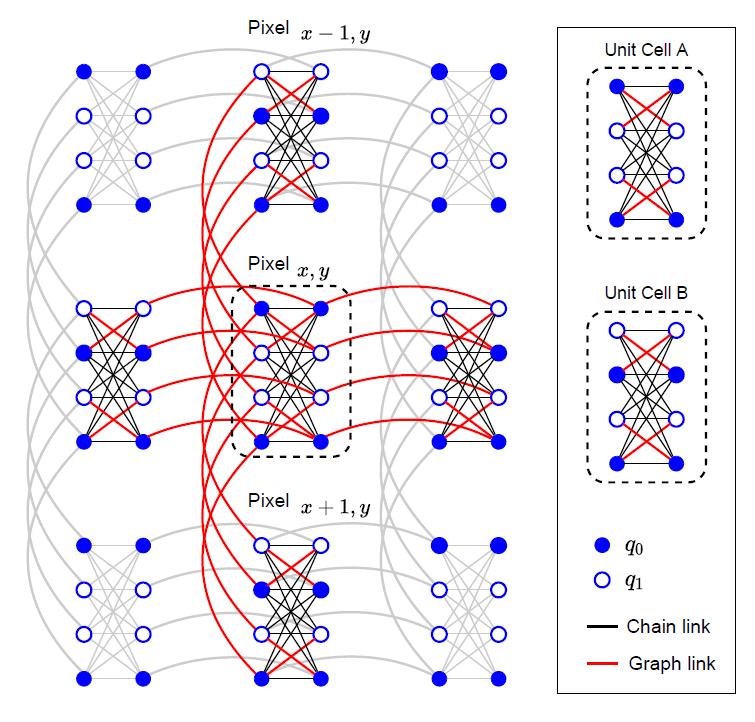}
\caption{The manual embedding mapping onto the Chimera graph \cite{kelany2022quantum}.}
\label{fig:cellTypeChimera}
\end{figure}

 Moreover, to provide necessary connections between adjacent pixels, we changed the location of rows corresponding to the qubits of MSB and LSB for neighboring cells as shown in Figure \ref{fig:cellTypeChimera}. Therefore, we introduced two types of cells. In \enquote{type A} cells, qubits of LSB are placed in the first and fourth row of a unit cell and qubits of the MSB are placed in the middle rows, while in \enquote{type B} cells, the second and the fourth row are considered for the LSB, and the remaining rows are associated with the MSB. Chimera cells are alternating between \enquote{type A} and \enquote{type B} in columns and in rows. This symmetric arrangement provides us with all necessary connections between binary variables. In figure \ref{fig:cellTypeChimera}, Red links show connections between different binary variables \cite{kelany2022quantum}.

\subsection{Mapping onto the Pegasus graph}\label{sec:mapPegasusGraph}
\begin{figure}[b]
\centering
\includegraphics[width=.35\textwidth]{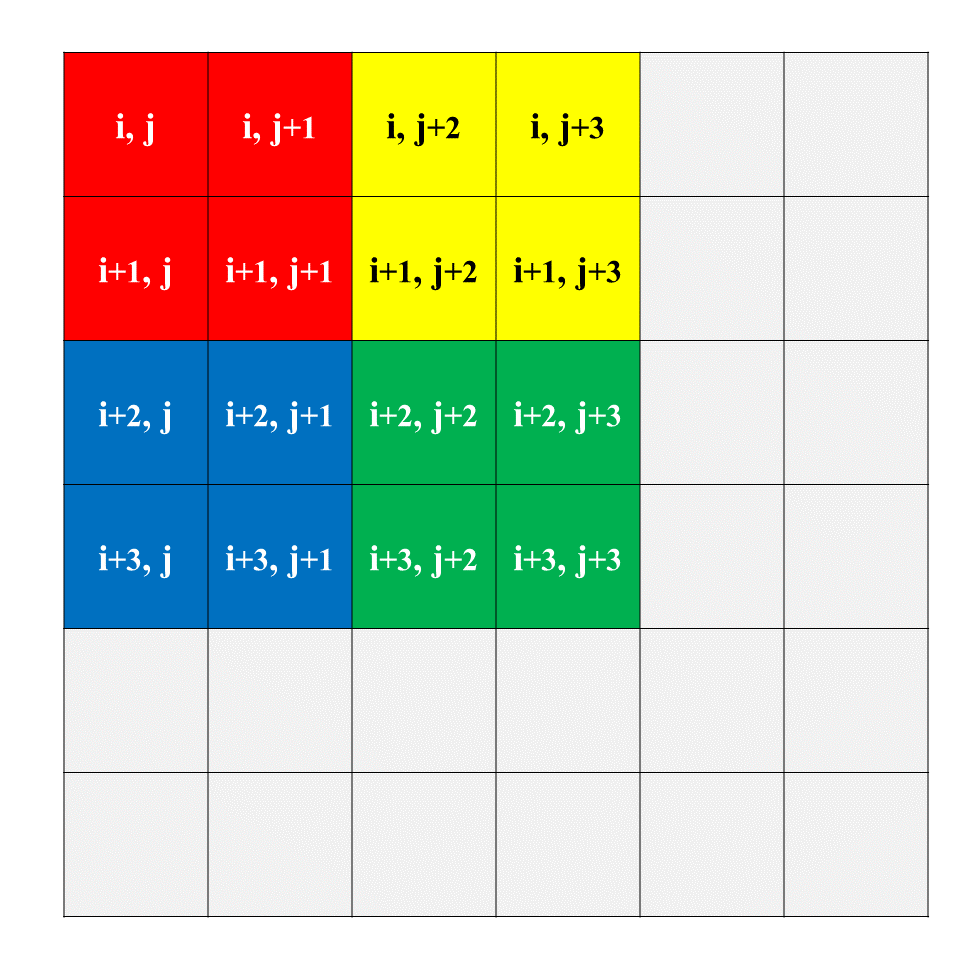}
\caption{Sub-images of size $2\times2$ shown in different colors.}
\label{fig:subimages}
\end{figure}

\begin{figure}
\centering
\includegraphics[width=.45\textwidth]{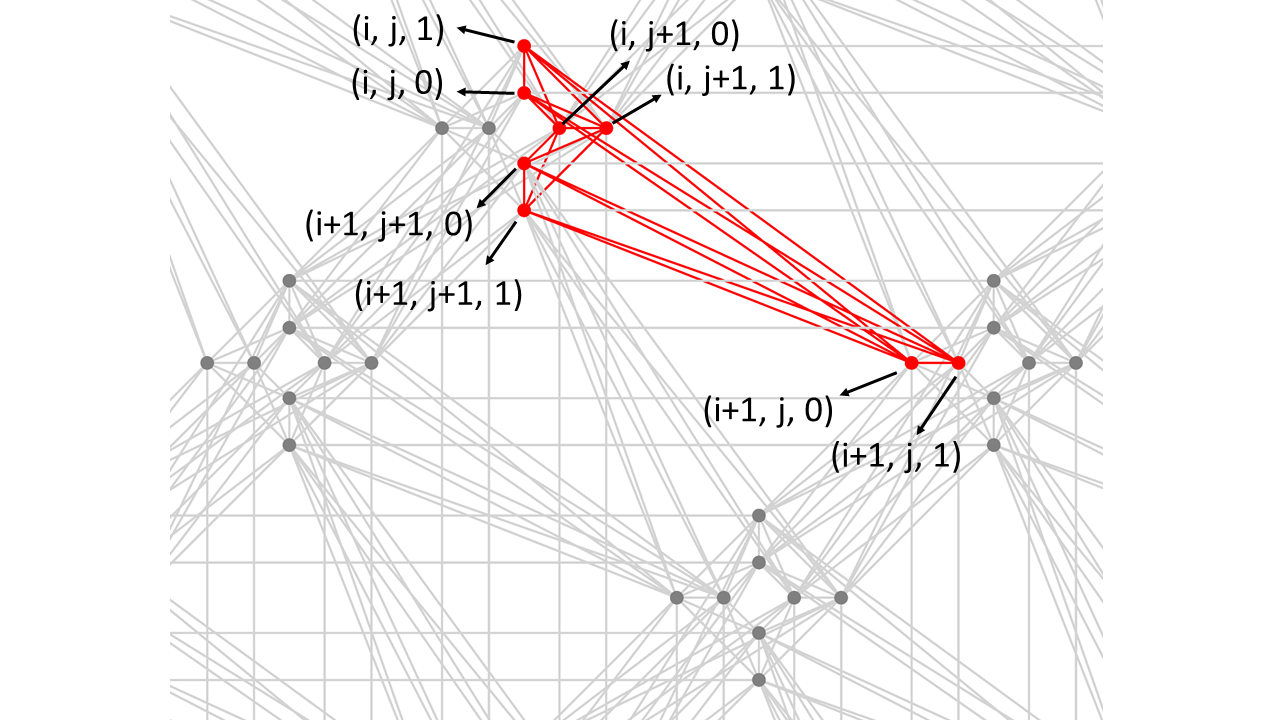}[t]
\caption{Mapping a $2\times2$ sub-image onto the Pegasus graph.}
\label{fig:phase1}
\end{figure}

To map the problem graph onto the Pegasus graph, we start with a sub-image of size $2\times2$ and then continue mapping its adjacent sub-images until all pixels are mapped. Consider an image with sub-images like Figure \ref{fig:subimages}. We start by mapping the red sub-image. This sub-image is mapped to eight qubits of two Chimera unit cells: six qubits of one Chimera unit cell, and two qubits of another one with the relative positions as Figure \ref{fig:phase1}.  
 
In a sub-image of size $2\times2$, each pixel has two adjacent pixels, and consequently, its binary variables should be connected to those of adjacent pixels. In the mapping of sub-images, these connections are provided. For example, binary variables of pixel $(i,j)$ in Figure \ref{fig:subimages} should be connected to those of $(i,j+1)$, and $(i+1,j)$ inside the red sub-image, and in the mapping of this sub-image, these connections exist. 
\par

All sub-images of size $2\times2$ can be mapped using the mentioned mapping. Pixels of a sub-image have connections with those of adjacent sub-images, and these connections should be provided in the mapping. We continue by mapping the right sub-image of the mapped one which is the yellow sub-image in Figure \ref{fig:subimages}. We can map this sub-image similar to what we did for the red one. However, the left pixels of this sub-image are connected to the right pixels of the red one (i.e., pixel $(i,j+2)$ is connected to pixel $(i,j+1)$, and pixel $(i+1,j+2)$ is connected to pixel $(i+1,j+1)$).
\par
\begin{figure}
\centering
\includegraphics[width=.4\textwidth]{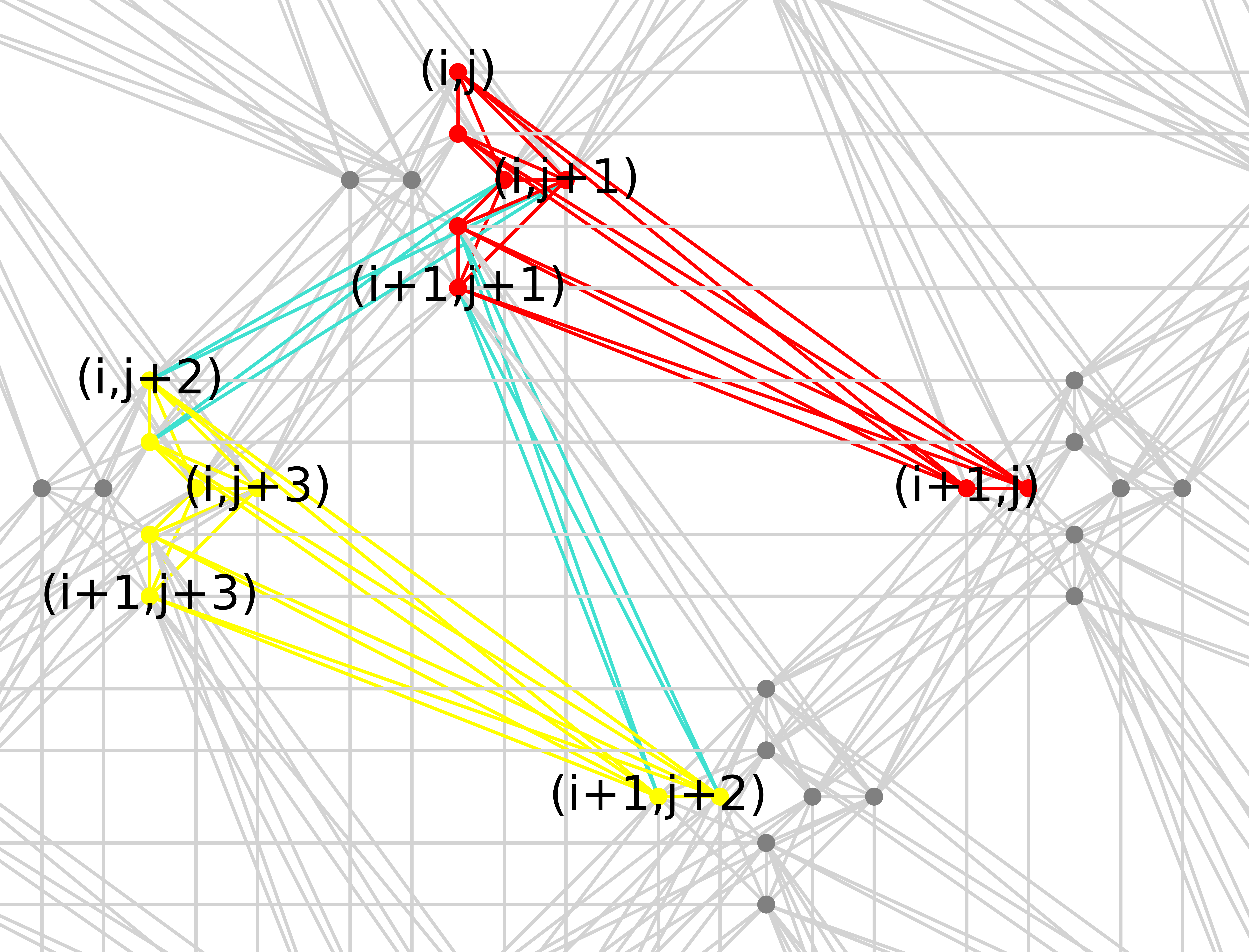}
\caption{Mapping of two horizontally adjacent sub-images onto the Pegasus graph. Red and yellow edges represent connections between pixels inside the red and the yellow sub-images respectively, while turquoise edges show the connections between adjacent pixels of those two sub-images.}
\label{fig:phase2}
\end{figure}

Consequently, we map the yellow sub-image in a specific relative location with the red sub-image to provide these connections. This is done by mapping the yellow sub-image in the lower left side of the red one as Figure \ref{fig:phase2}. In this figure, red and yellow edges show connections inside the red and the yellow sub-images respectively, while the turquoise edges show the connections between pixels of the red sub-image and pixels of the yellow sub-image. To avoid congestion in the plots, the third coordinate representing the bit of each pixel is excluded in drawing the labels.
\newline
We can map every other two horizontally adjacent sub-images with the same approach. 

\par
\begin{figure}
\centering
\includegraphics[width=.44\textwidth]{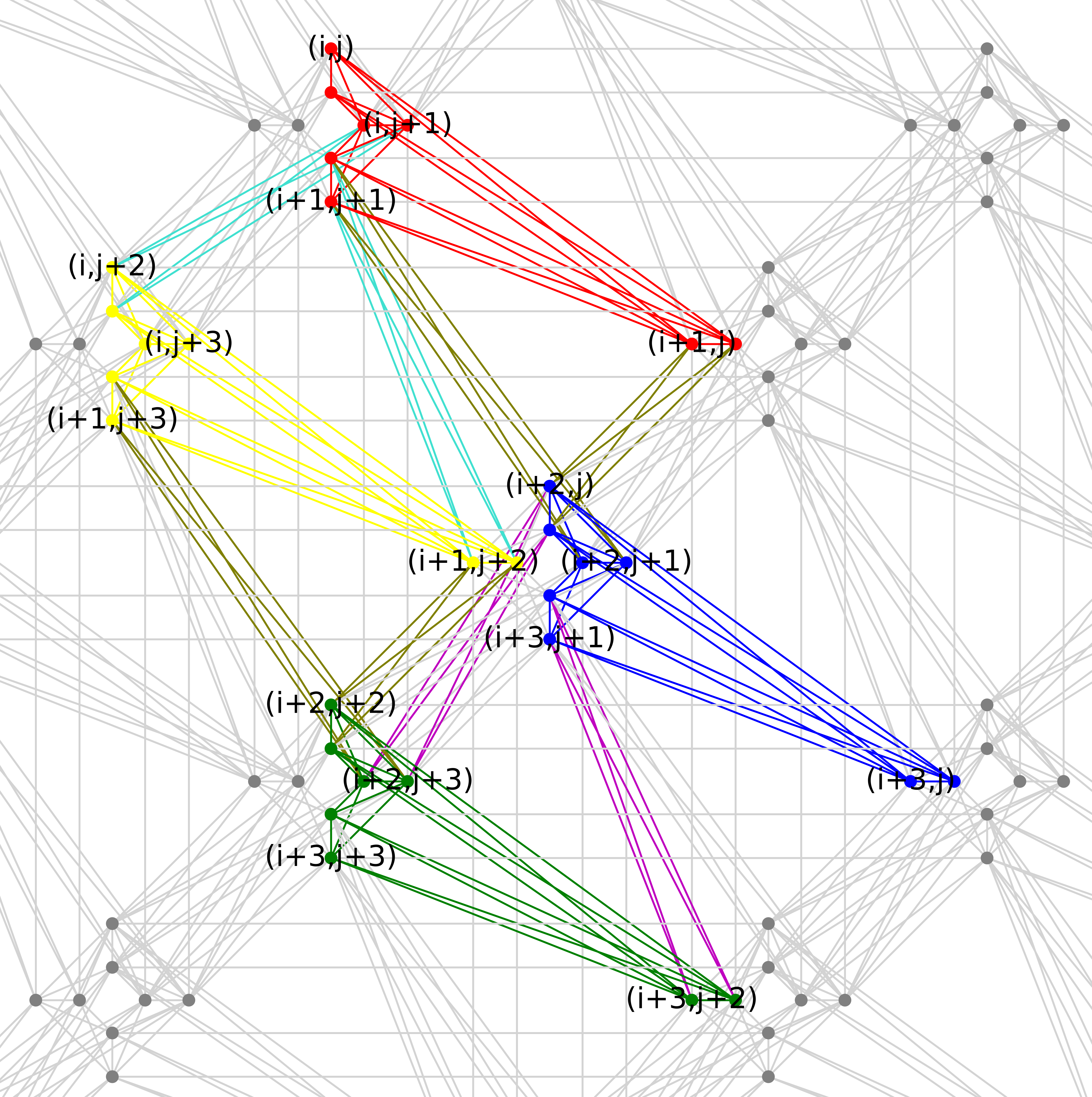}
\caption{Mapping four adjacent sub-images onto the Pegaus graph. Connections between the blue and green sub-images are shown in magenta while connections between the red and blue sub-images as well as connections between the yellow and green sub-images are shown in olive.}
\label{fig:phase3}
\end{figure}

Now that we mapped two sub-images, we continue by mapping the button sub-images of those that were mapped (the blue and green sub-images in Figure \ref{fig:subimages}). These two sub-images are also horizontally adjacent and we can map them the same as the previous adjacent sub-images. However, the top pixels of these two sub-images have connections with the bottom pixels of the previous sub-images (i.e., pixel $(i+2,j)$ is connected to pixel $(i+1,j)$, pixel $(i+2,j+1)$ is connected to pixel $(i+1,j+1)$, and so on). We map these two sub-images with a relative location with the previous ones such that these connections are provided. This is accomplished by mapping them in the lower right side of the previously mapped sub-images as Figure \ref{fig:phase3}. 
\par

In Figure \ref{fig:phase3}, the pair of blue and green sub-images are mapped similarly to the pair of red and yellow sub-images. Connections between blue and green sub-images are shown in magenta. These edges are like the turquoise edges as both are of the horizontally adjacent sub-images. Also, connections between red and blue sub-images as well as connections between the yellow and green sub-images are shown in olive. Red and blue sub-images are vertically adjacent and so are yellow and green ones. As a result, connections between red and blue sub-images are similar to those between yellow and green ones.

A Chimera cell that was partially used in the mapping of previous sub-images now is completely used as the mapping of new sub-images matched the previous one.

\par
Four sub-images are mapped so far and we can continue the same procedure to map another four sub-images. The location of the next four sub-images follows the approach that we used to map adjacent sub-images. I.e., the right four sub-images of the mapped one are mapped in the lower left direction of them similar to what we did for mapping the yellow sub-image after the red one. Also, the bottom four sub-images of the mapped one can be mapped in the lower right direction of them as we did for mapping blue and green sub-images after red and yellow sub-images. Continuing this trend, we can map the whole image.
\par
In the end, the location of pixels after mapping is congruent with the location of them in the original ones. However, they rotated by 45 degrees counterclockwise and then reflected over the y-axis. Except for marginal sub-images, all other Chimera unit cells are fully used for mapping as the mapping of sub-images matches together and forms a consistent graph of Chimera unit cells.

\section{Experiments}\label{sec:experiments}
In this section, we present experimental performance results of three different types of embeddings, i.e., Pegasus Native embedding (the one that we proposed in this paper), Chimera symmetrical embedding deployed originally on the Chimera network of the $2000Q\_6$ machine and ten automatically generated embeddings using D-Wave's Ocean tool. Table I shows the properties of these embeddings. Five Automatic embeddings ($A1$, $A2$, ..., and $A5$) are generated for the Advantage machine and another five ($A6$, $A7$, ..., and $A10$) for the $2000Q_6$ machine. 
\par

\subsection{Setup}
\subsubsection{Terminology}
RCLO: The Ratio of Chain-Length-One (RCLO) is the proportion of chains of length one in the population of chains in the embedding. It provides an additional measure of how compact is the embedding.  

\subsubsection{Configurations}\label{paragragh:configuration}
We used the default amounts for the annealing parameters in our experiments. I.e., annealing\_time$=20\mu s$ and number\_of\_reads $= 1000$.

\subsubsection{Datasets}
The datasets consist of simulated (SAR) data with a medium noise level (SNR=10dB) and medium complexity (Perlin correlation=18). This dataset presents a large exploration space, with a wide spectrum that includes high-frequency data that present a challenge for the phase unwrapping process.
The dataset includes $5$ synthetic images with the size of $10\times10$ pixel, generated using a Perlin Noise Generator \cite{perlin1985image}. 

\begin{table}[b]
\caption{Embedding Properties}

\begin{center}
\begin{tabular}{|l|c|c|c|c|}
\hline
\textbf{Embedding}&\multicolumn{4}{|c|}{\textbf{Chain Length}} \\
\cline{2-5} 
\textbf{Type} & \textbf{\textit{Avg}}& \textbf{\textit{STD}}& \textbf{\textit{RCLO}}& \textbf{\textit{Max}} \\
\hline
Pegasus native & 1 & 0 & 1 & 1\\
Chimera symmetric & 4 & 0 & 0 & 4\\
A$^{\mathrm{*}}$1 (Advantage) & 1.890 & 1.024 & 0.455 & 5\\
A2 (Advantage) & 1.595 & 0.782 & 0.580 & 4\\
A3 (Advantage) & 1.830 & 0.825 & 0.430 & 4\\
A4 (Advantage) & 1.685 & 0.804 & 0.510 & 4\\
A5 (Advantage) & 1.795 & 0.783 & 0.405 & 4\\
A6 (2000Q\_6) & 4.145 & 1.321 & 0 & 9\\
A7 (2000Q\_6) & 6.365 & 4.152 & 0.005 & 29\\
A8 (2000Q\_6) & 4.640 & 1.955 & 0.005 & 15\\
A9 (2000Q\_6) & 4.155 & 1.171 & 0 & 7\\
A10 (2000Q\_6) & 4.870 & 1.762 & 0 & 11\\
\hline
\multicolumn{4}{l}{$^{\mathrm{*}}$Automatic}
\end{tabular}
\par\end{center}
\label{tbl:Embedding properties.}
\end{table}

\subsubsection{Accuracy Metrics}\label{paragragh:Accuracy Metrics}
To determine how close two images (of identical size) are to each other, we use the \emph{matching fraction} metric defined as the fraction of pixels that are identical in the two images.
To evaluate the accuracy of our methods, we compare the obtained image to
Noisy Unwrapped Ground-truth images which are the images obtained by a sensor or synthetically by adding noise to synthetic noise-free ground-truth images\cite{kelany2022quantum}.

\subsection{Results}
In terms of chain length, our proposed Pegasus native embedding is the optimum embedding with the lowest possible chain length. D-Wave's Ocean software couldn't find any other embedding with an average chain length of one or even close to one (Table 1). Furthermore, for the $2000Q\_6$ machine, our proposed Chimera symmetric embedding in \cite{kelany2022quantum} has the lowest chain length in comparison with the other five embeddings generated by D-Wave. As shown in Table 1, average and maximum chain lengths of embeddings for the Advantage machine are lower than those of the $2000Q\_6$ machine as the Pegasus architecture has more connectivities than the Chimera architecture. 
\par
The obtained accuracy for images of the dataset using seven embeddings (the Pegasus native, the Chimera symmetric, and five Automatics)  for the Advantage machine is reported in Table \ref{tbl:Accuracy}. The accuracy of equal to or more than $98\%$ for all images and the average accuracy of $99\%$ was obtained from our proposed Pegasus native embedding, and it outperforms other embeddings by far. 
\par
\begin{figure}
\centering
\includegraphics[width=.35\textwidth]{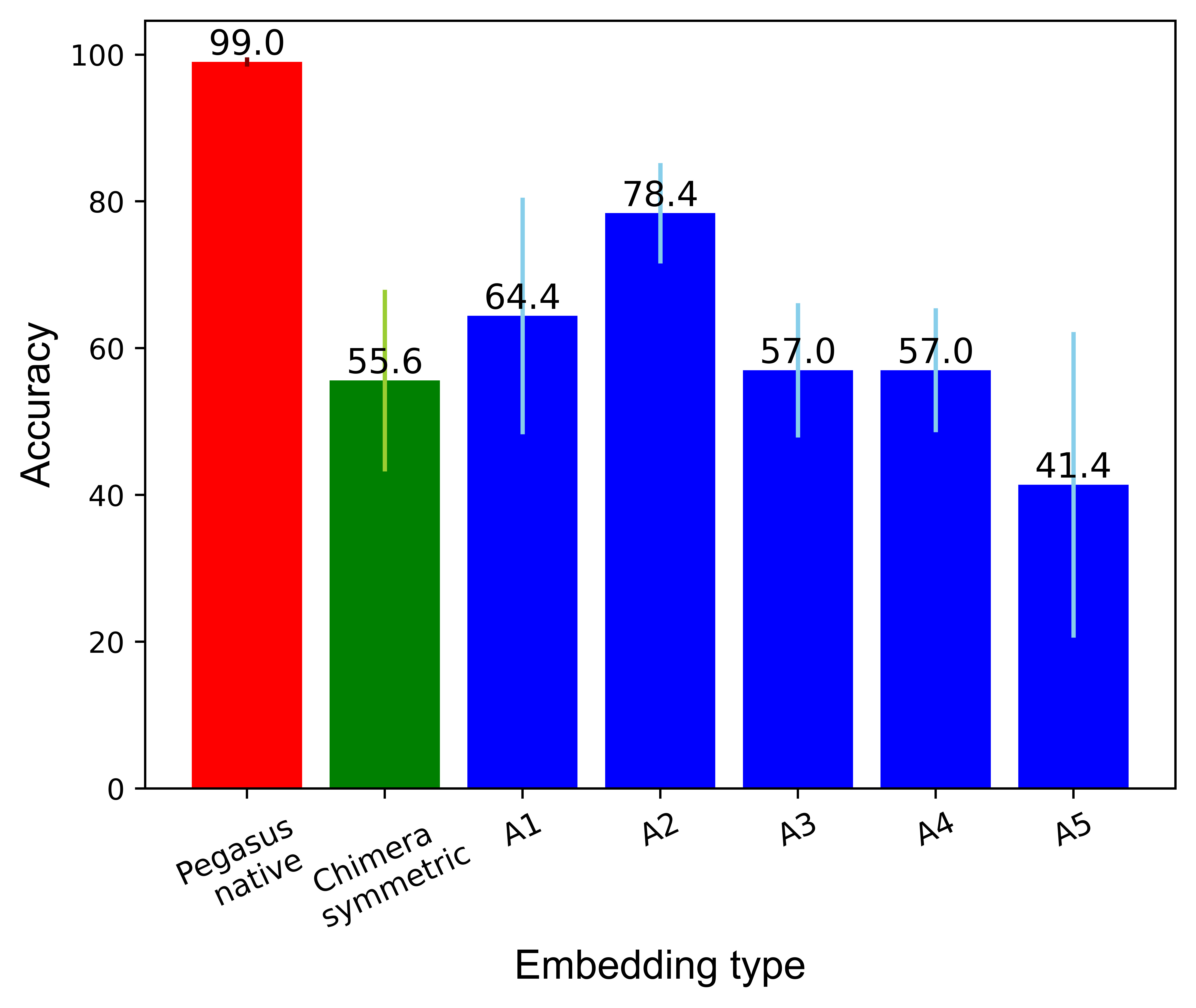}
\caption{Average accuracy of different embeddings on the Advantage machine. Red, green, and blue columns correspond to our Pegasus native, Chimera symmetric, and Automatic embeddings respectively. The error bars show the standard deviation.}
\label{fig:averageaccuracyadv}
\end{figure}
\begin{figure}[b]
\centering
\includegraphics[width=.35\textwidth]{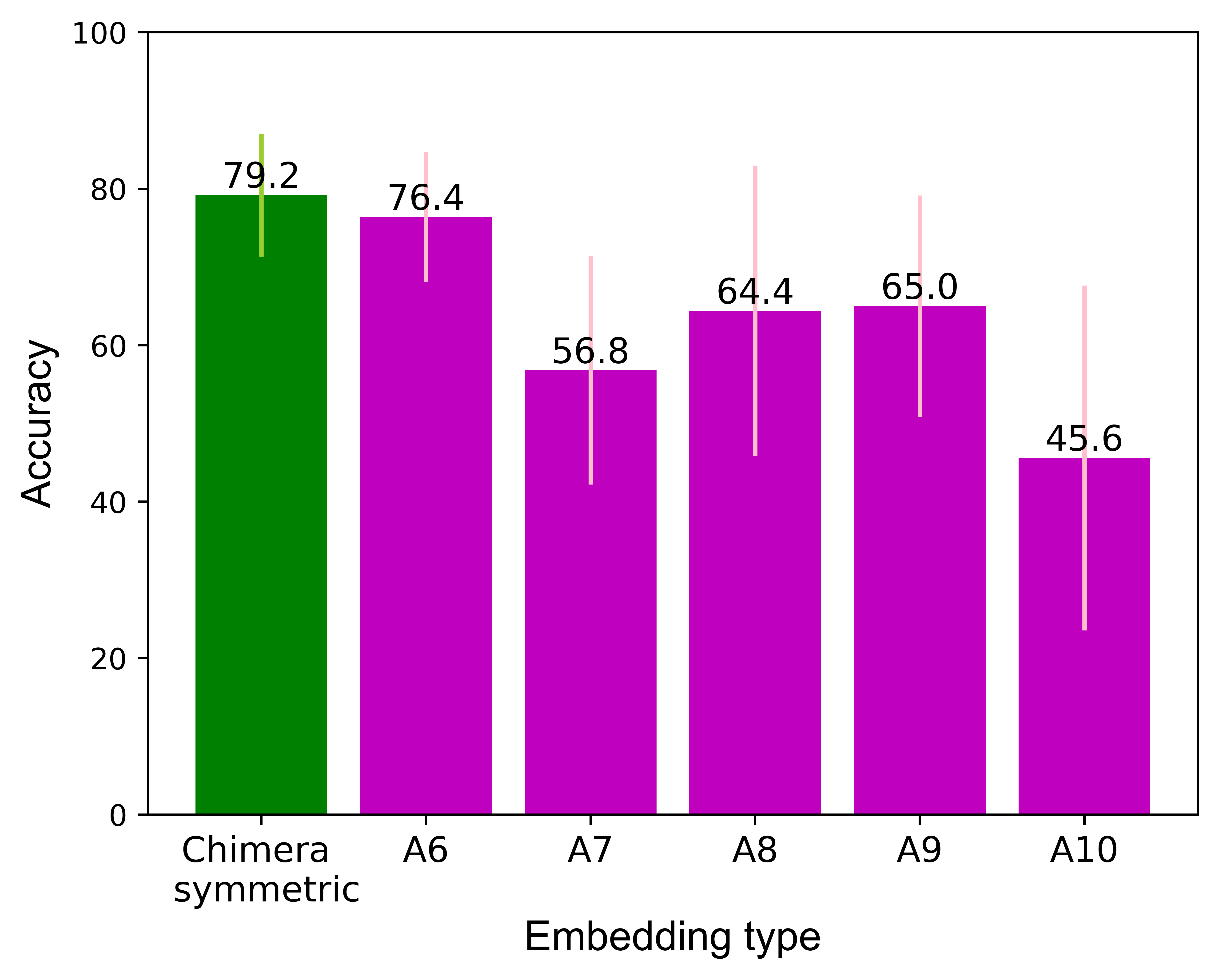}
\caption{Average accuracy of different embeddings on $2000Q\_6$ machine. Green and purple columns correspond to the Chimera symmetric and Automatic embeddings respectively. The error bars show the standard deviation.}
\label{fig:averageaccuracy2000}
\end{figure}

In the Advantage machine, the Chimera symmetric embedding has a performance similar to Automatic embeddings. The average obtained accuracy for all Automatic embeddings is $59.64\%$, and slightly better than Chimera symmetric embedding. Figure \ref{fig:averageaccuracyadv} shows the average accuracy of all embeddings in the Advantage machine. However, for the $2000Q\_6$ machine, as illustrated in Figure \ref{fig:averageaccuracy2000}, the Chimera symmetric embedding outperforms Automatic embeddings. The obtained results for $2000Q\_6$ machine are reported in Table \ref{tbl:Accuracyp}.

\begin{table}[b]
\caption{Obtained accuracy using different embeddings on the Advantage\_system6.1 machine}
\begin{center}
\begin{tabular}{|l|c|c|c|c|c|c|c|}
\hline
\textbf{Image}&\multicolumn{7}{|c|}{\textbf{Embedding Type}} \\
\cline{2-8} 
\textbf{number}&\parbox{1cm}{\centering \textbf{Pegasus} \\[-0.6ex] \textbf{native}} & \parbox{1.2cm}{\centering \textbf{Chimera} \\[-0.6ex] \textbf{symmetric}}& \textbf{\textit{A1}}& \textbf{\textit{A2}}& \textbf{\textit{A3}}& \textbf{\textit{A4}}&\textbf{\textit{A5}}  \\
\hline
Image 1 & 99 & 46 & 75 & 66 & 67 & 54 & 29\\
Image 2 & 100 & 40 & 34 & 85 & 44 & 44 & 19\\
Image 3 & 99 & 71 & 69 & 84 & 59 & 57 & 26\\
Image 4 & 98 & 69 & 80 & 80 & 49 & 70 & 69\\
Image 5 & 99 & 52 & 64 & 77 & 66 & 60 & 64\\
\cline{1-8} 
Average & 99 & 55.6 & 64.4 & 78.4 & 57 & 57 & 41.4\\
\hline
\end{tabular}
\par\end{center}
\label{tbl:Accuracy}
\end{table}

\begin{table}
\caption{Obtained Accuracy using different embeddings on the $2000Q\_6$ machine}
\begin{center}
\begin{tabular}{|l|c|c|c|c|c|c|}
\hline
\textbf{Image}&\multicolumn{6}{|c|}{\textbf{Embedding type}} \\
\cline{2-7} 
\textbf{number} & \parbox{1.2cm}{\centering \textbf{Chimera} \\[-0.6ex] \textbf{symmetric}}& \textbf{\textit{A6}}& \textbf{\textit{A7}}& \textbf{\textit{A8}}& \textbf{\textit{A9}}&\textbf{\textit{A10}}  \\
\hline
Image 1 & 86 & 75 & 71 & 84 & 64 & 33  \\
Image 2 & 64 & 62 & 34 & 62 & 43 & 14 \\
Image 3 & 82 & 86 & 62 & 72 & 77 & 58 \\
Image 4 & 84 & 83 & 71 & 74 & 83 & 79 \\
Image 5 & 80 & 76 & 46 & 30 & 58 & 44 \\
\cline{1-7} 
Average & 79.2 & 76.4 & 56.8 & 64.4 & 65 & 45.6 \\
\hline
\end{tabular}
\par\end{center}
\label{tbl:Accuracyp}
\end{table}

Our experiments seem to indicate that short chain lengths result in better performance (accuracy). Thus the Pegasus native and the Chimera symmetric, having the shortest chains in the respective architectures, resulted in the best performance for our phase unwrapping problem.
\newline
Similarly, from the automatic embeddings derived by D-WAVE’s Ocean tool, the ones with the shortest average-length chains (A2 and A6) had the best performance. However, the performance is not linearly related to the average chain length. 
\par
As examples to the contrary, one can consider mappings A5 and A1 where A5 has a shorter average chain length (1.795 vs. 1.895 for A1 ) yet the performance of the A5 embedding is worse than that of the A1 (41.4 vs. 64.4 for A1). Similar behavior can be seen for A10 as compared to A7.
\par

Finally, the Chimera symmetric embedding was used in both the Advantage (Pegasus) and the $2000Q$ (Chimera) machines. Given that Chimera is a subgraph of Pegasus, one would have expected similar performance obtained from both architectures. Yet, the $2000Q$ machine yielded consistently better performance than the Advantage one. This result is consistent with the results reported in \cite{willsch2022benchmarking} and needs further investigation.
\par
The above observation notwithstanding, the Advantage architecture outperforms the 2000Q as its increased degree of connectivity results in much shorter chains.

\section{Conclusion}\label{sec:conclusion}
In this work, we proposed a heuristic mapping to embed the phase unwrapping problem onto the D-Wave's Advantage architecture. This embedding is the optimal one in terms of chain length. Automatic embeddings generated by D-Wave's Ocean tool couldn't find any embedding with an average chain length close to ours. We experimentally showed that our embedding outperforms others significantly. Our experiments also confirmed that lower average chain length of embedding on D-Wave's machines could generally result in a better performance for the phase unwrapping problem.

\clearpage

\appendices

\section{Cost Derivation}
\label{FirstAppendix}
Denoting by $\varphi_{i}$ the phase of pixel $i$, and by $\phi_{i}$
the wrapped phase of the same pixel, we can relate the phase and wrapped
phases of pixels $i$ and $j$ as follows.

\begin{equation}
\varphi_{i}=\phi_{i}+2\pi k_{i}
\label{eq:a1}
\end{equation}

and

\begin{equation}
\varphi_{j}=\phi_{j}+2\pi k_{j}
\label{eq:a2}
\end{equation}

Further, due to the Nyquist criterion, and if pixels $i$ and $j$ are neighbouring, then

\begin{equation}
\mid \varphi_{i}-\varphi_{j} \mid <\pi\,.
\label{eq:Nyquist}
\end{equation}
or
\begin{equation}
-\pi < \varphi_{i}-\varphi_{j} < \pi
\end{equation}

and using \eqref{eq:a1} and \eqref{eq:a2}
\begin{equation}
-\pi < \phi_{i}-\phi_{j}  + 2\pi (k_i -k_j) < \pi
\end{equation}
or
\begin{equation}
-\frac{1}{2} < \frac{\phi_i -\phi_j}{2\pi} + (k_i -k_j) <\frac{1}{2}
\end{equation}
Denoting the nearest integer (or round) function as $nint(.)$ 
then
\begin{equation}
nint(\frac{\phi_i -\phi_j}{2\pi} + (k_i -k_j))=0
\end{equation}
 Since $k_i - k_j$ is an integer, then 

\begin{dmath*}
nint(\frac{\phi_i -\phi_j}{2\pi} + (k_i -k_j))= 
\end{dmath*}
\begin{dmath}
nint(\frac{\phi_i -\phi_j}{2\pi} )+ (k_i -k_j)=0
\label{eq:a4}
\end{dmath}

denoting 
\begin{equation}
a_{ij} \stackrel{\text{def}}{=}-nint(\frac{\phi_i -\phi_j}{2\pi} )
\end{equation}

then equation \eqref{eq:a4} can be rewritten as

\begin{equation}
k_{i}-k_{j}-a_{ij}=0 \Rightarrow k_{i}-k_{j}=a_{ij}
\end{equation}

\vspace{0.8em}

This equation is the basis of the cost function the optimization of
which will produce appropriate values for the labels $k_{i}$.

The unwrapping problem can then be expressed as an optimization problem
of the cost function 
\begin{equation}
E=\sum_{(s,t)\text{\ensuremath{\in}}A}W_{st}|k_{t}-k_{s}-a_{st}|\,,
\end{equation}
that is, 
\begin{equation}
\arg\min_{k}\left[\sum_{(s,t)\text{\ensuremath{\in}}A}W_{st}|k_{t}-k_{s}-a_{st}|\right]\,,
\end{equation}
where $k_{i}$ are the labels that will determine the original phase
as per Equation (1), $A$ is the set of pixels in the SAR image, and $W_{st}$
are weights defining the neighborhood structure.

\section*{Acknowledgment}
This research was supported in part by grants from the Natural Sciences and Engineering Research Council of Canada (NSERC) through its Discovery grants program and by Quantum BC.

\par

\bibliographystyle{ieeetr}
\bibliography{citations}

\end{document}